\documentclass[amsmath,floatfix,twocolumn,superscriptaddress]{revtex4-1}
\usepackage{subfigure}
\usepackage{siunitx}
\usepackage{amssymb}
\usepackage{amsmath}
\usepackage{graphicx}
\usepackage{array}
\usepackage{dcolumn}
\usepackage{psfrag}
\usepackage{bm}
\usepackage{color}
\usepackage{multirow}

\begin{document}

\title{Crystal Structure Prediction and Phase Stability in Highly Anharmonic Silver-Based Chalcohalide Anti-Perovskites}

\author{Pol Benítez}
    \affiliation{Group of Characterization of Materials, Departament de F\'{i}sica, Universitat Polit\`{e}cnica de Catalunya,
    Campus Diagonal-Bes\`{o}s, Av. Eduard Maristany 10--14, 08019 Barcelona, Spain}
    \affiliation{Research Center in Multiscale Science and Engineering, Universitat Politècnica de Catalunya,
    Campus Diagonal-Bes\`{o}s, Av. Eduard Maristany 10--14, 08019 Barcelona, Spain}

\author{Cibrán López}
    \affiliation{Group of Characterization of Materials, Departament de F\'{i}sica, Universitat Polit\`{e}cnica de Catalunya,
    Campus Diagonal-Bes\`{o}s, Av. Eduard Maristany 10--14, 08019 Barcelona, Spain}
    \affiliation{Research Center in Multiscale Science and Engineering, Universitat Politècnica de Catalunya,
    Campus Diagonal-Bes\`{o}s, Av. Eduard Maristany 10--14, 08019 Barcelona, Spain}

\author{Cong Liu}
    \affiliation{Departament de F\'{i}sica, Universitat Polit\`{e}cnica de Catalunya, Campus Nord, Jordi Girona 1--3, 
    08005 Barcelona, Spain}

\author{Ivan Caño}
    \affiliation{Research Center in Multiscale Science and Engineering, Universitat Politècnica de Catalunya,
    Campus Diagonal-Bes\`{o}s, Av. Eduard Maristany 10--14, 08019 Barcelona, Spain}
    \affiliation{Department of Electronic Engineering, Universitat Politècnica de Catalunya, 08034 Barcelona, Spain}

\author{Josep-Llu\'is Tamarit}
    \affiliation{Group of Characterization of Materials, Departament de F\'{i}sica, Universitat Polit\`{e}cnica de Catalunya,
    Campus Diagonal-Bes\`{o}s, Av. Eduard Maristany 10--14, 08019 Barcelona, Spain}
    \affiliation{Research Center in Multiscale Science and Engineering, Universitat Politècnica de Catalunya,
    Campus Diagonal-Bes\`{o}s, Av. Eduard Maristany 10--14, 08019 Barcelona, Spain}

\author{Edgardo Saucedo}
    \affiliation{Research Center in Multiscale Science and Engineering, Universitat Politècnica de Catalunya,
    Campus Diagonal-Bes\`{o}s, Av. Eduard Maristany 10--14, 08019 Barcelona, Spain}
    \affiliation{Department of Electronic Engineering, Universitat Politècnica de Catalunya, 08034 Barcelona, Spain}

\author{Claudio Cazorla}
    \affiliation{Group of Characterization of Materials, Departament de F\'{i}sica, Universitat Polit\`{e}cnica de Catalunya,
    Campus Diagonal-Bes\`{o}s, Av. Eduard Maristany 10--14, 08019 Barcelona, Spain}
    \affiliation{Research Center in Multiscale Science and Engineering, Universitat Politècnica de Catalunya,
    Campus Diagonal-Bes\`{o}s, Av. Eduard Maristany 10--14, 08019 Barcelona, Spain}

\begin{abstract}
	Silver-based chalcohalide anti-perovskites (CAP), Ag$_{3}$BC (B = S, Se; C = Cl, Br, I), represent an emerging 
	family of energy materials with intriguing optoelectronic, vibrational and ionic transport properties. However, 
	the structural features and phase stability of CAP remain poorly investigated to date, hindering their fundamental 
	understanding and potential integration into technological applications. Here we employ theoretical first-principles 
	methods based on density functional theory to fill this knowledge gap. Through crystal structure prediction 
	techniques, \textit{ab initio} molecular dynamics simulations, and quasi-harmonic free energy calculations, we 
	unveil a series of previously overlooked energetically competitive phases and temperature-induced phase transitions 
	for all CAP. Specifically, we identify a new cubic $P2_{1}3$ structure as the stable phase of all CAP containing 
	S both at zero temperature and $T \neq 0$~K conditions. Consequently, our calculations suggest that the cubic 
	$Pm\overline{3}m$ phase identified in room-temperature X-ray diffraction experiments is likely to be metastable. 
	Furthermore, for CAP containing Se, we propose different orthorhombic ($Pca2_{1}$ and $P2_{1}2_{1}2_{1}$) and 
	cubic ($I2_{1}3$) structures as the ground-state phases and reveal several phase transformations induced by 
	temperature. This theoretical investigation not only identifies new candidate ground-state phases and solid-solid 
	phase transformations for all CAP but also provides insights into potential stability issues affecting technological
	applications based on these highly anharmonic materials. 
\end{abstract}

\maketitle

\section{Introduction}
\label{sec:intro}
Silver-based chalcohalide anti-perovskites (CAP) with chemical formula Ag$_{3}$BC (B = S, Se; C = Cl, Br, I) are 
structurally similar to the lead halide perovskites (e.g., CsPbI$_{3}$), with the ``anti'' designation indicating the 
exchange of anions and cations compared to the typical ionic perovskite arrangement. Analogous to lead halide perovksites, 
CAP are highly promising materials for energy and optoelectronic applications, boasting low toxicity due to their lead-free 
composition \cite{palazon22,ghorpade23}. The two most extensively studied CAP compounds, Ag$_{3}$SBr and Ag$_{3}$SI, 
possess experimental band gaps of approximately $1.0$~eV \cite{luna23,cano24}, rendering them favorable for photovoltaic 
applications. These have been also recognized as room-temperature superionic conductors \cite{takahashi66,hull04}. 
Additionally, CAP have been investigated as potential thermoelectric materials \cite{kawamura80,magistris72} owing to their 
substantial vibrational anharmonicity and intriguing transport properties \cite{wakamura90,sakuma85}. 

The current surge in interest in CAP is evident from the recent publication of several experimental works detailing 
improved synthesis methods. The pioneering synthesis of the archetypal CAP Ag$_{3}$SI and Ag$_{3}$SBr dates back to 
$1960$, credited to Reuter and Hardel \cite{reuter60}. However, the majority of early CAP synthesis techniques required 
vacuum conditions, elevated temperatures ($> 600$~K), and prolonged synthesis durations, often spanning during days or 
even months. More recently, rapid synthesis routes have emerged, enabling the production of high-purity Ag$_{3}$SI and 
Ag$_{3}$SBr powders at temperatures below $600$~K through mechanochemical methods \cite{luna23}. Furthermore, advancements 
have also led to the synthesis of CAP solid solutions at moderate temperatures ($< 500$~K), such as Ag$_{3}$SBr$_{x}$I$_{1-x}$ 
($0 \le x \le 1$), using a cost-effective solution-processing approach \cite{cano24}.

\begin{figure*}
    \centering
    \includegraphics[width=0.9\linewidth]{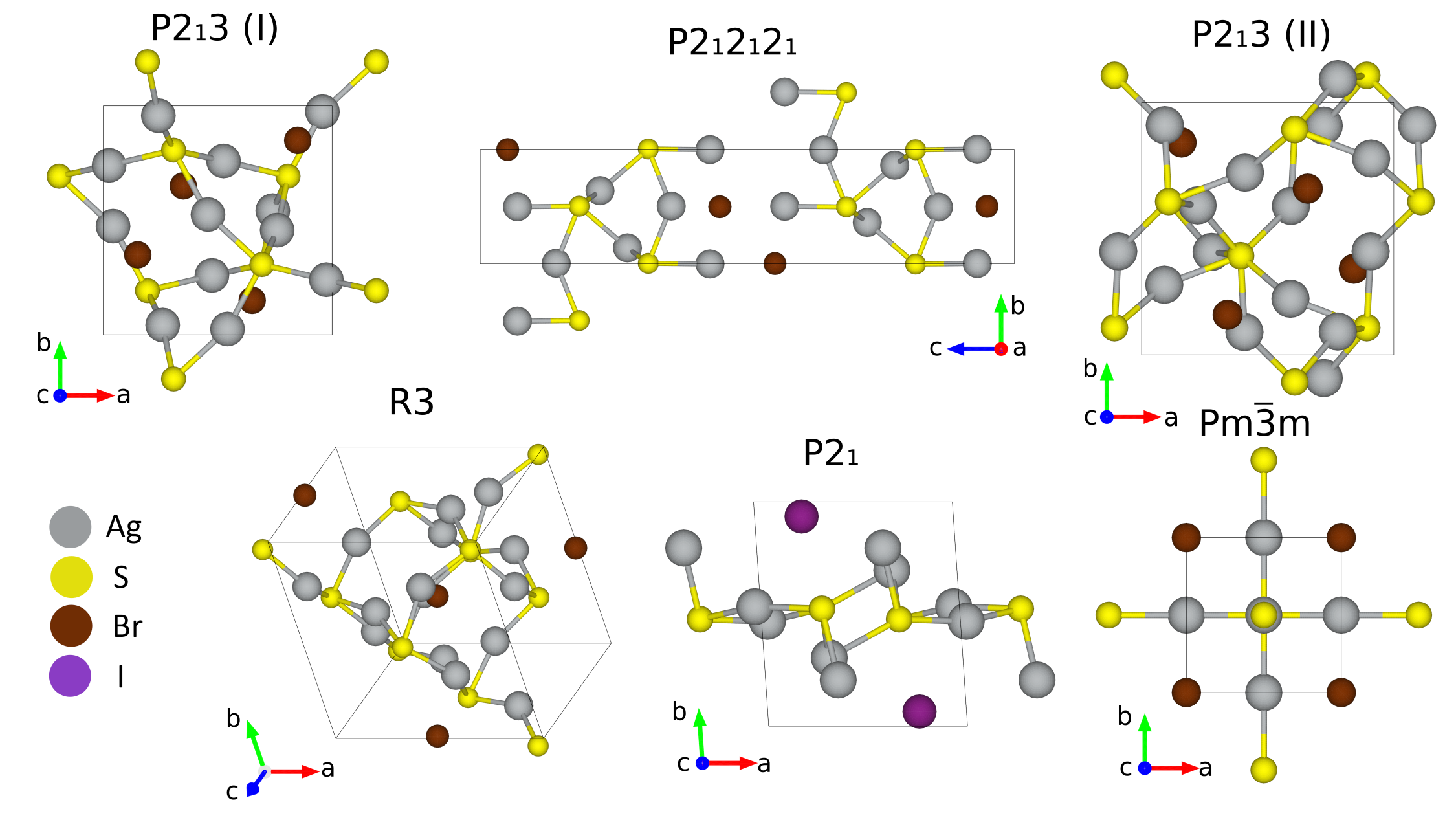}
        \caption{\textbf{Sketch of some energetically competitive crystal structures predicted with CSP-DFT methods 
        for the archetypal silver-based chalcohalide anti-perovskites Ag$_{3}$SBr and Ag$_{3}$SI.} The cubic 
	$Pm\overline{3}m$ phase that has been experimentally identified as the corresponding room-temperature $\beta$ 
	phase is shown for comparison. Additional representations of other energetically competitive crystal structures 
	are provided in the Supplementary Fig.1.}
    \label{fig1}
\end{figure*}

Despite these recent synthesis advances, the phase stability and structural properties of CAP remain poorly 
investigated and understood to date. For both Ag$_{3}$SBr and Ag$_{3}$SI, three different polymorphs have been consistently 
reported: a low-temperature $\gamma$ phase ($< 130$--$160$~K), a room-temperature phase denoted as $\beta$ with low ionic 
conductivity, and an $\alpha$ phase at high temperatures ($> 500$~K) exhibiting high ionic Ag$^{+}$ conductivity and 
S$^{-}$/I$^{-}$ or S$^{-}$/Br$^{-}$ chemical disorder \cite{hoshino78,sakuma80,hoshino81}. Regarding Ag$_{3}$SCl and the rest 
of CAP compounds containing Se, to the best of our knowledge, there is no experimental data available concerning their structural 
and phase stability properties.

In the original reports by Hoshino \emph{et al.} \cite{hoshino78,sakuma80,hoshino81}, the low-temperature $\gamma$ phase was 
described as trigonal with space group $R3$ for Ag$_{3}$SI and orthorhombic with space group $Cmcm$ for Ag$_{3}$SBr. The 
room-temperature $\beta$ phase for both Ag$_{3}$SI and Ag$_{3}$SBr was identified as cubic with space group $Pm\overline{3}m$
and the high-temperature disordered $\alpha$ phase as cubic with space group $Im\overline{3}m$ \cite{hull04,cho94,yin20}. 
In addition, a metastable room-temperature phase exhibiting high ionic conductivity, denoted as $\alpha^{*}$ and 
obtained through quenching of the $\alpha$ phase, has been reported for Ag$_{3}$SI \cite{hull04,cho94}. However, subsequent 
studies on Ag$_{3}$SBr suggested that the crystal symmetry of the low-temperature $\gamma$ phase could be more precisely 
described as monoclinic or triclinic, rather than orthorhombic \cite{honda07}. Furthermore, investigations into several 
Ag$_{3}$SBr$_{x}$I$_{1-x}$ solid solutions revealed that for a range of mixed compositions the low-temperature $\gamma$ 
phase was orthorhombic with space group $Pnm2_{1}$ or $Pnmn$ \cite{honda07}.

Since all existing CAP synthesis methods involve temperatures well above ambient conditions, temperatures at which Ag$_{3}$SI  
and Ag$_{3}$SBr exhibit ionic conductivity, it is very likely that upon annealing the samples remain dynamically arrested in 
metastable states characterized by ionic disorder. This phenomenon is originated by the existence of significant energy barriers 
that hinder the transition towards energetically more favorable ordered phases. A similar behavior has been recently 
demonstrated for halide hybrid perovskites \cite{shahrokhi22}. This probable tendency poses a significant challenge in 
identifying truly stable low-temperature phases in CAP. Systematic studies able to precisely evaluate the relative stability 
of different CAP polymorphs are therefore urgently needed. These investigations are crucial for improving our fundamental 
understanding of CAP and to properly assess their potential for possible technological applications. 

In this study, we employ theoretical first-principles methods based on density functional theory to discern 
candidate stable phases for all CAP compounds at both zero-temperature and $T \neq 0$ conditions. Notably, the 
predicted lowest-energy CAP phases, systematically identified through crystal structure prediction, \textit{ab initio}
molecular dynamics, and quasi-harmonic free energy techniques, differ from those previously reported in experimental 
studies. Therefore, besides providing original and valuable data for Ag$_{3}$SCl and other CAP compounds containing Se, 
the present study prompts a reassessment of the prevalent phase diagrams for Ag$_{3}$SI and Ag$_{3}$SBr, urging 
further experimental investigations into CAP. 

\begin{figure*}
    \centering
    \includegraphics[width=\linewidth]{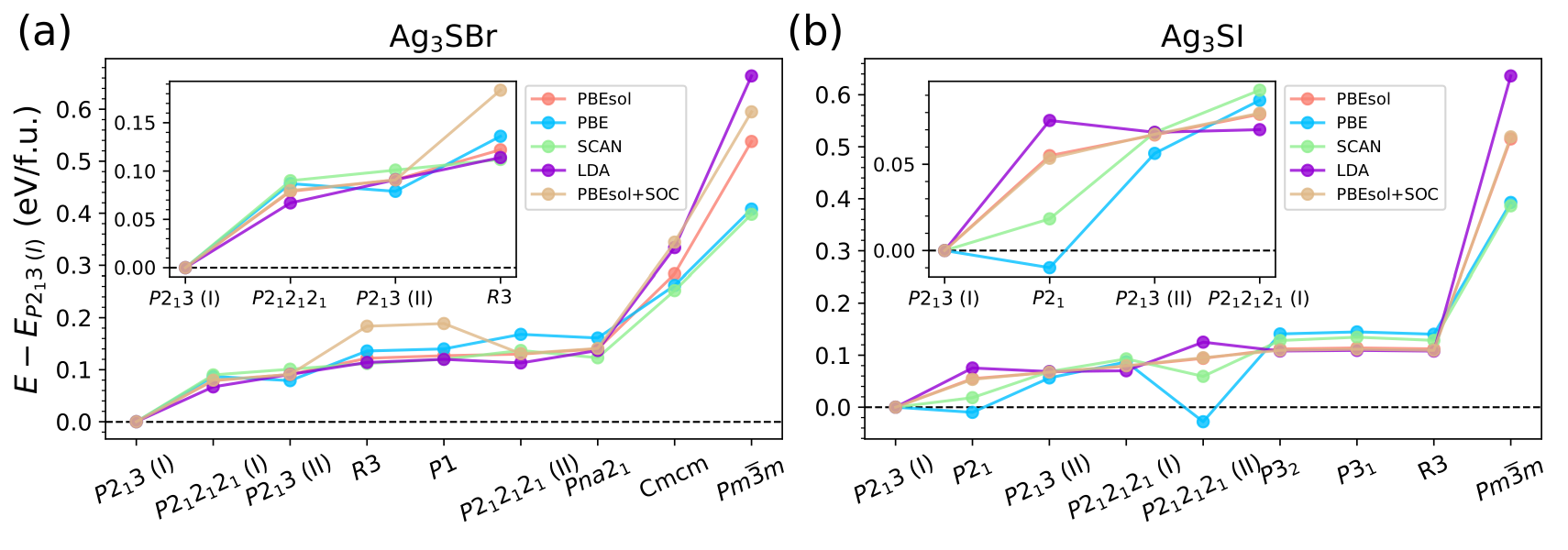}
	\caption{\textbf{Energy of competitive crystal structures calculated for archetypal CAP using different 
	DFT exchange-correlation functionals.} (a)~Ag$_{3}$SBr and (b)~Ag$_{3}$SI. The energy of the cubic 
	$Pm\overline{3}m$ phase that is experimentally resolved at room temperature is shown for comparison.}
    \label{fig2}
\end{figure*}

\section{Results}
\label{sec:results}

\begin{figure*}
    \centering
    \includegraphics[width=\linewidth]{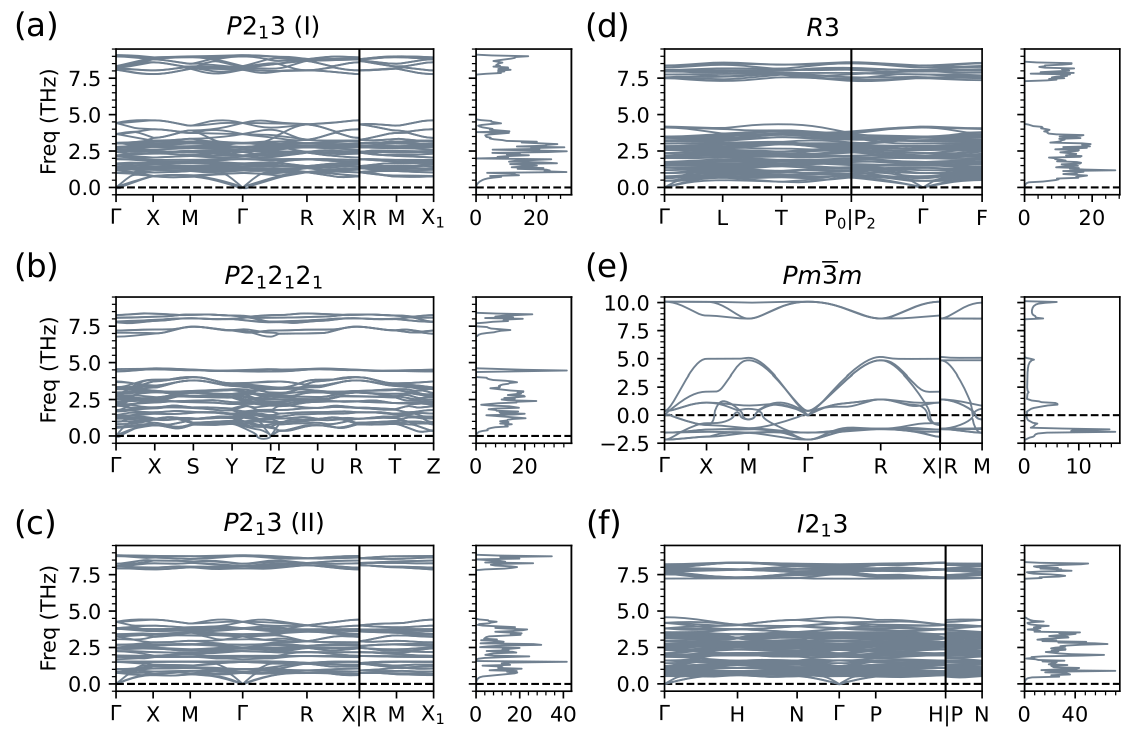}
        \caption{\textbf{Vibrational phonon spectrum of Ag$_{3}$SBr calculated for different crystal structures.}
        Phonon calculations were performed at zero-temperature conditions. All considered phases are vibrationally
        stable made the exception of the cubic $Pm\overline{3}m$ phase, which exhibits abundant imaginary phonon
        frequencies and branches. Results were obtained with the semi-local functional PBEsol \cite{pbesol}.}
    \label{fig3}
\end{figure*}

\subsection{Zero-temperature crystal structure prediction}
\label{subsec:zero}
We conducted zero-temperature crystal structure prediction (CSP) calculations for the two archetypal CAP compounds,
Ag$_{3}$SBr and Ag$_{3}$SI, using the MAGUS software \cite{magus} and considering a maximum of $20$ atoms per unit
cell (Methods). MAGUS employs an evolutionary algorithm augmented with machine learning techniques and graph theory. 
The initial structures proposed by MAGUS were subsequently relaxed using first-principles methods based on density 
functional theory (DFT, Methods). Figure~\ref{fig1} illustrates some of the energetically most favorable phases 
identified in the rankings resulting from our CSP-DFT calculations (additional structural representations can be 
found in Supplementary Fig.1). Notable among them are two distinct cubic structures with space group $P2_{1}3$, an 
orthorhombic $P2_{1}2_{1}2_{1}$ phase, a monoclinic $P2_{1}$ phase and a trigonal $R3$ phase. 

\begin{figure*}
    \centering
    \includegraphics[width=\linewidth]{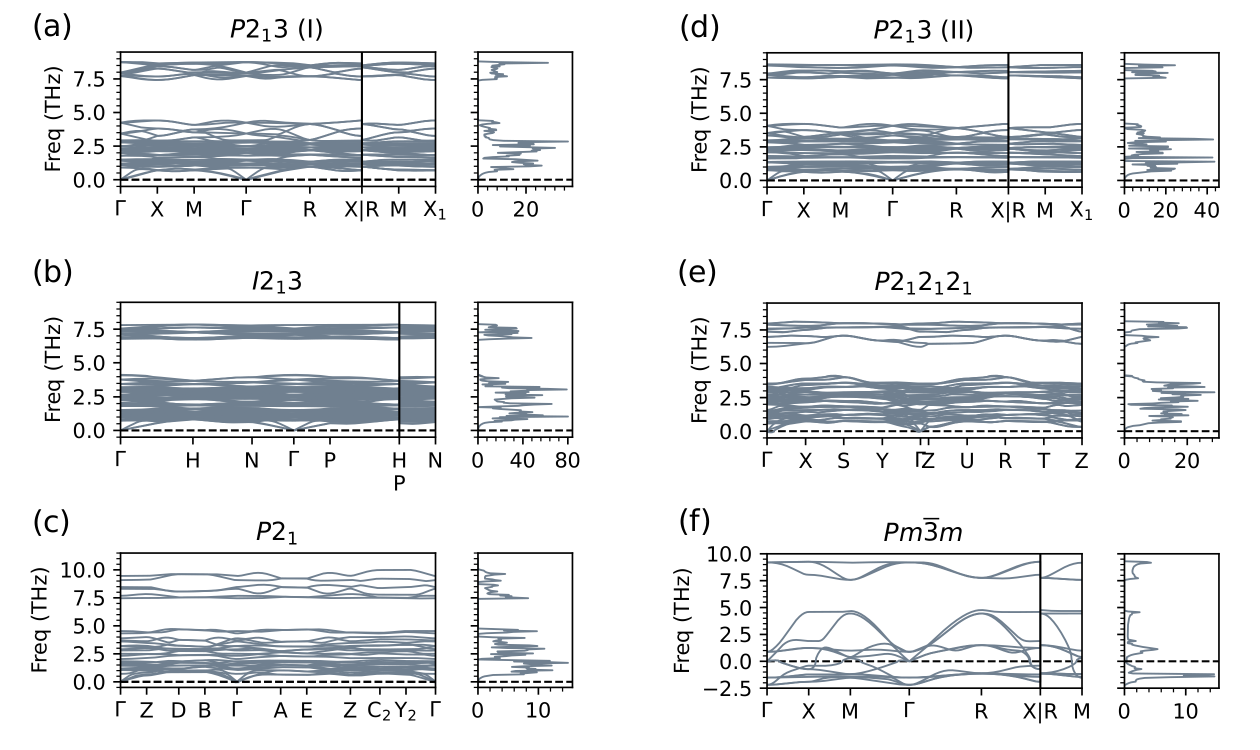}
        \caption{\textbf{Vibrational phonon spectrum of Ag$_{3}$SI calculated for different crystal structures.}
        Phonon calculations were performed at zero-temperature conditions. All considered phases are vibrationally
        stable made the exception of the cubic $Pm\overline{3}m$ phase, which exhibits abundant imaginary phonon
        frequencies and branches. Results were obtained with the semi-local functional PBEsol \cite{pbesol}.}
    \label{fig4}
\end{figure*}

Figure~\ref{fig2} presents the polymorph energy rankings obtained for Ag$_{3}$SBr and Ag$_{3}$SI using various 
DFT exchange-correlation energy ($E_{xc}$) functionals, including LDA \cite{lda}, PBE \cite{pbe}, PBEsol \cite{pbesol} and 
SCAN \cite{scan}. This set of DFT functionals includes different local and semilocal approximations to $E_{xc}$, which 
must be assessed to identify possible limitations of the employed DFT methods \cite{angelis18}. For Ag$_{3}$SBr, all 
the examined DFT functionals consistently identify the cubic $P2_{1}3$~(I) structure as the ground-state phase (Fig.\ref{fig2}a). 
At $T = 0$~K conditions, the second energetically most favorable structure is an orthorhombic $P2_{1}2_{1}2_{1}$ phase, followed 
by a cubic $P2_{1}3$~(II) and a trigonal $R3$ phase. Notably, the PBE functional is the sole functional predicting the cubic 
$P2_{1}3$~(II) phase to be marginally more favorable than the orthorhombic $P2_{1}2_{1}2_{1}$ phase. Incorporating spin-orbit 
coupling (SOC) effects in the DFT calculations appears inconsequential for determining the most energetically competitive phases, 
as indicated by the similarity between the PBEsol and PBEsol+SOC curves. 

Remarkably, our crystal CSP-DFT calculations do not classify the orthorhombic $Cmcm$ or the cubic $Pm\overline{3}m$ 
phases as energetically competitive, despite their proposal as the $\gamma$ and $\beta$ phases of Ag$_{3}$SBr from 
experiments \cite{hoshino78,sakuma80,hoshino81}. Consistently, across all the examined DFT functionals, the energy of 
these two phases is several hundreds of meV per formula unit (f.u.) higher than that of the theoretical ground state, 
the cubic $P2_{1}3$~(I) phase (Fig.\ref{fig2}a).

For Ag$_{3}$SI, the energy ranking is also dominated by the cubic $P2_{1}3$~(I) phase, closely followed by a monoclinic
$P2_{1}$ phase (Fig.\ref{fig2}b). However, the PBE functional deviates from the other energy functionals, suggesting 
an orthorhombic $P2_{1}2_{1}2_{1}$ phase as the ground state. In this case, accounting for SOC effects in the 
DFT calculations neither poses a significant variation in the energy difference results, as evidenced by the nearly 
identical curves for PBEsol and PBEsol+SOC. Another cubic $P2_{1}3$~(II) phase emerges as energetically competitive 
across all surveyed DFT functionals. 

Likewise to Ag$_{3}$SBr, the energy of the experimentally observed low-temperature trigonal $R3$ and room-temperature 
cubic $Pm\overline{3}m$ phases of Ag$_{3}$SI is estimated to be substantially higher than that of the corresponding 
ground-state phase ($> 0.1$~eV/f.u.) by all the examined functionals. It is worth noting the similarity in energy 
among the trigonal $P3_{2}$ and $P3_{1}$ phases identified in our CSP-DFT calculations and the experimentally 
identified trigonal $R3$ phase, which stems from their structural resemblance.   

Figures~\ref{fig3}--\ref{fig4} show the vibrational phonon spectra calculated for the four most energetically 
favorable structures determined for Ag$_{3}$SBr and Ag$_{3}$SI at zero temperature, respectively, along with those 
of the cubic $Pm\overline{3}m$ phase and another cubic $I2_{1}3$ phase introduced in the next section. Across 
the four predicted stable polymorphs, we observe no imaginary phonon frequencies along their high-symmetry reciprocal 
space paths, indicating their vibrational stability. Interestingly, all these phases exhibit a wide frequency band 
gap spanning approximately from $5.0$ to $7.5$~THz, with the exception of the orthorhombic $P2_{1}2_{1}2_{1}$ phase
that displays slightly lower frequencies in the high-frequency regime. 

Conversely, the cubic $Pm\overline{3}m$ phase, proposed as the room-temperature $\beta$ phase of both Ag$_{3}$SBr 
and Ag$_{3}$SI \cite{sakuma80,hoshino81}, exhibits numerous imaginary phonon frequencies along all high-symmetry 
reciprocal space paths, indicating its vibrational instability at $T = 0$~K conditions. A possible $T$-induced 
stabilization of this cubic $Pm\overline{3}m$ phase will be discussed further in Sec.\ref{sec:discussion}. Additionally, 
the orthorhombic $Cmcm$ phase, suggested as the low-temperature $\gamma$ phase of Ag$_{3}$SBr \cite{sakuma80,hoshino81}, 
exhibits also abundant imaginary phonon frequencies (Supplementary Fig.2). Consequently, this phase is vibrationally 
unstable at low temperatures and, even ignoring its highly unfavourable energy, cannot be regarded as the ground-state 
phase. On the contrary, the trigonal $R3$ phase, proposed as the low-temperature $\gamma$ phase of Ag$_{3}$SI 
\cite{sakuma80,hoshino81}, appears to be vibrationally stable at $T = 0$~K conditions (Supplementary Fig.2); 
however, in view of its high energy (Fig.\ref{fig2}b), this phase should be regarded as metastable in the low-temperature 
regime. 

For completeness, we calculated the electronic band-structure properties of Ag$_{3}$SBr and Ag$_{3}$SI in the 
predicted ground state ($P2_{1}3$) and experimentally observed room-temperature phase ($Pm\overline{3}m$), 
as shown in Supplementary Fig.3. These electronic band-structure calculations were performed using the hybrid HSE06 
functional \cite{hse06}, including spin-orbit coupling effects (HSE06+SOC), and based on the equilibrium structures 
obtained at the PBEsol level \cite{pbesol}. In all cases, the estimated band gaps were identified as indirect, with 
values ranging from $1.4$ to $1.9$~eV (Supplementary Fig.3). Ag$_{3}$SBr in the $Pm\overline{3}m$ phase exhibited 
the largest band gap, while Ag$_{3}$SI in the $P2_{1}3$ phase presented the smallest. 

As demonstrated in this section, our CSP-DFT calculations yield a significantly different set of candidate ground-state 
phases compared to those experimentally proposed for Ag$_{3}$SBr and Ag$_{3}$SI. Specifically, a cubic $P2_{1}3$ phase 
consistently emerges as the most energetically favorable phase at low temperatures in our calculations, closely followed 
by an orthorhombic $P2_{1}2_{1}2_{1}$ and a monoclinic $P2_{1}$ phase, depending on the material. Importanly, these 
new candidate ground-state phases are all shown to be vibrationally stable, contrasting with, for example, the 
orthorhombic $Cmcm$ phase proposed as the lowest energy structure of Ag$_{3}$SBr. Consequently, our theoretical findings 
call for a reevaluation of previous experimental characterizations of CAP conducted at low temperatures.

\begin{figure}
    \centering
    \includegraphics[width=\linewidth]{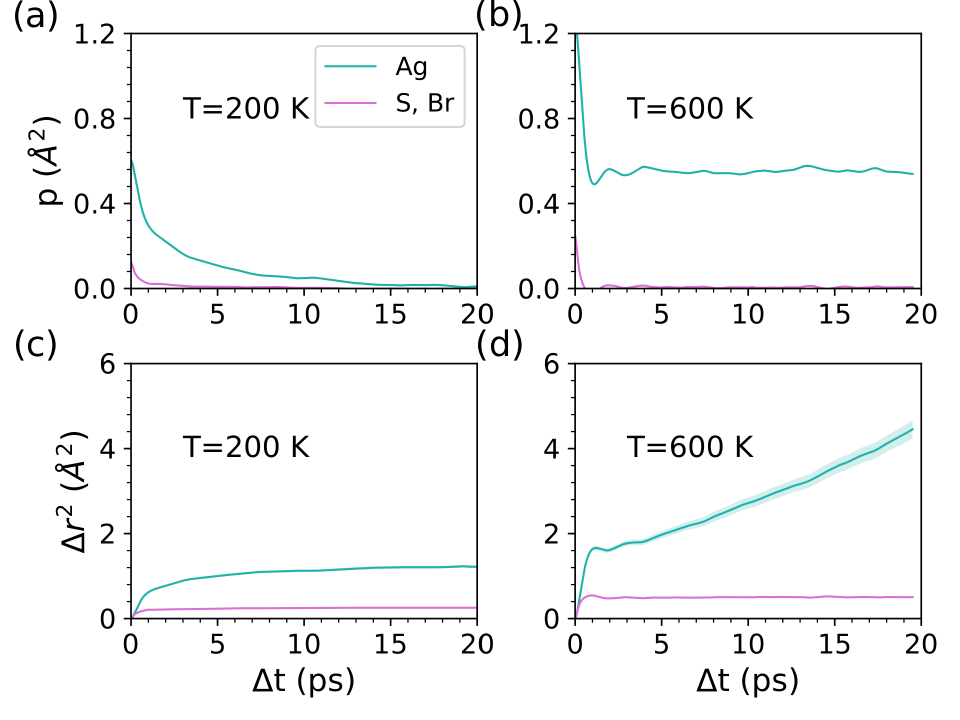}
	\caption{\textbf{Analysis of the vibrational stability and ionic diffusion of Ag$_{3}$SBr in the cubic
	$Pm\overline{3}m$ phase at $T \neq 0$ conditions.} (a)--(b)~Position correlation function \cite{cazorla12,cazorla14}, 
	$p (\Delta t)$, estimated at $T = 200$ and $600$~K. (c)--(d)~Mean square displacement, $\Delta r^{2} (\Delta t)$, 
	estimated at $T = 200$ and $600$~K. Results were obtained from \textit{ab initio} molecular dynamics simulations 
	(Methods) performed with the semi-local functional PBEsol \cite{pbesol}.}
    \label{fig5}
\end{figure}

\begin{figure}
    \centering
    \includegraphics[width=\linewidth]{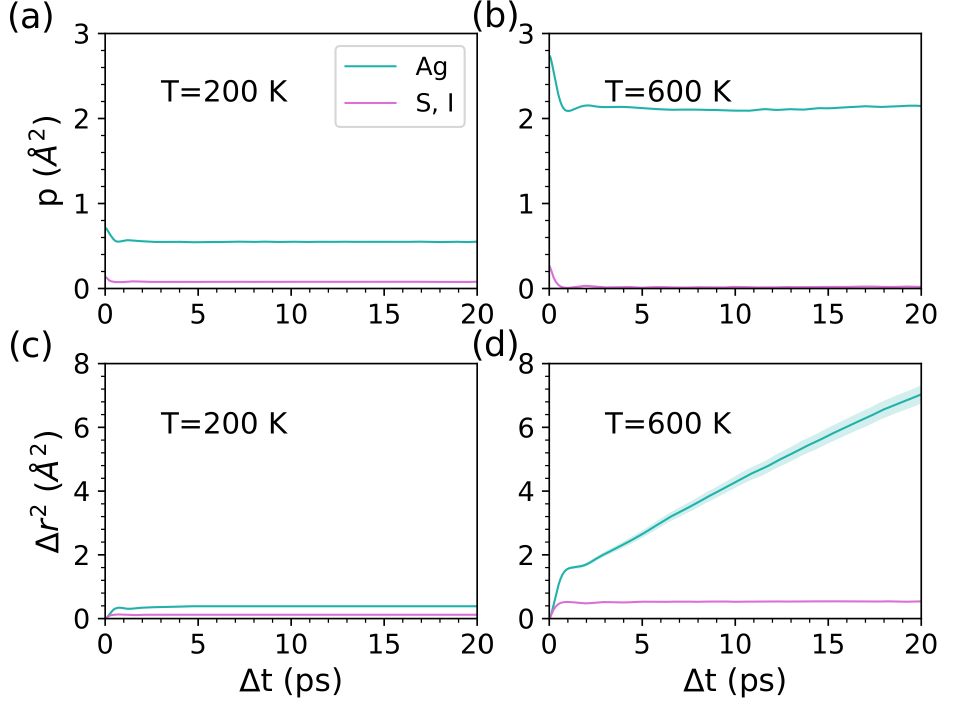}
	\caption{\textbf{Analysis of the vibrational stability and ionic diffusion of Ag$_{3}$SI in the cubic
	$Pm\overline{3}m$ phase at $T \neq 0$ conditions.} (a)--(b)~Position correlation function \cite{cazorla12,cazorla14}, 
	$p (\Delta t)$, estimated at $T = 200$ and $600$~K. (c)--(d)~Mean square displacement, $\Delta r^{2} (\Delta t)$,
	estimated at $T = 200$ and $600$~K. Results were obtained from \textit{ab initio} molecular dynamics simulations 
	(Methods) performed with the semi-local functional PBEsol \cite{pbesol}.}
    \label{fig6}
\end{figure}

\begin{figure*}
    \centering
    \includegraphics[width=\textwidth]{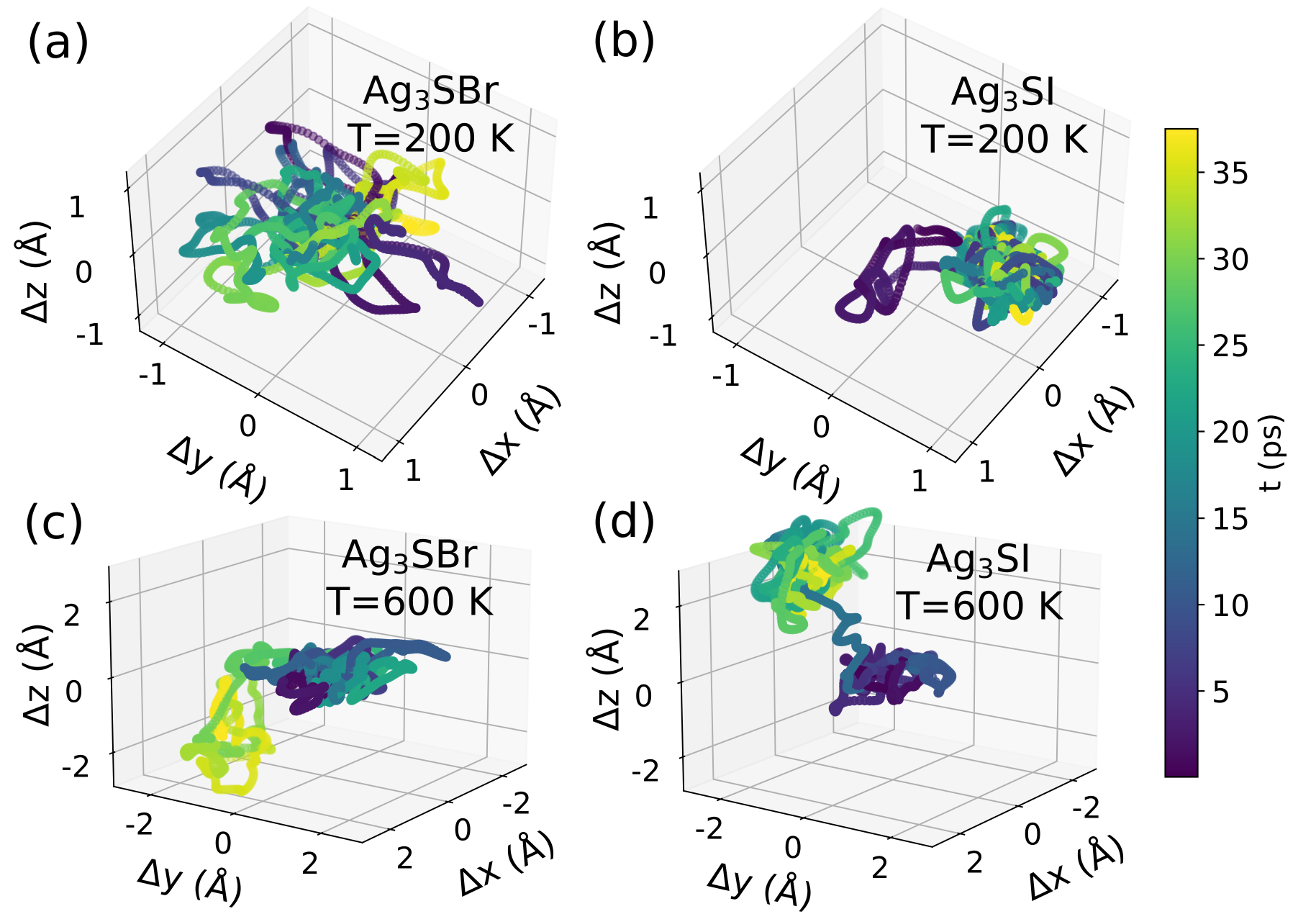}
	\caption{\textbf{Analysis of Ag ion trajectories obtained from \textit{ab initio} molecular dynamics 
	simulations.} Ag$_{3}$SBr in the cubic $Pm\overline{3}m$ phase at (a)~$T = 200$~K and (c)~$T = 600$~K.
	Ag$_{3}$SI in the cubic $Pm\overline{3}m$ phase at (b)~$T = 200$~K and (d)~$T = 600$~K. Results were
	obtained with the semi-local functional PBEsol \cite{pbesol}.}
    \label{fig7}
\end{figure*}

\begin{figure*}
    \centering
    \includegraphics[width=\linewidth]{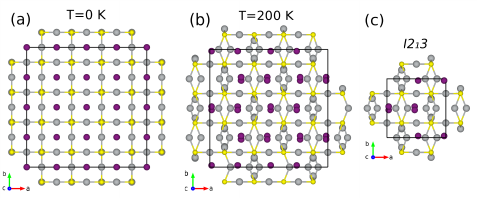}
        \caption{\textbf{Ag$_{3}$SI supercell representations obtained from \textit{ab initio} molecular dynamics 
        simulations.} (a)~The starting cubic $Pm\overline{3}m$ phase at zero temperature. (b)~The distorted cubic
        $Pm\overline{3}m$ phase found at $T = 200$~K. (c)~The new cubic $I2_{1}3$ phase determined from the distorted
        cubic $Pm\overline{3}m$ phase. Ag, S and I ions are represented with grey, yellow and purple spheres, 
	respectively. Results were obtained with the semi-local functional PBEsol \cite{pbesol}.}
    \label{fig8}
\end{figure*}

\subsection{Finite-temperature crystal structure prediction}
\label{subsec:finite}
The observed low-temperature vibrational instability of the cubic $Pm\overline{3}m$ phase (Figs.\ref{fig3}--\ref{fig4}),
proposed as the room-temperature $\beta$ phase of both Ag$_{3}$SBr and Ag$_{3}$SI \cite{sakuma80,hoshino81}, prompted us
to perform comprehensive \emph{ab initio} molecular dynamics (AIMD) simulations at $T \neq 0$~K conditions (Methods). The
primary objective of these dynamical simulations was to assess the vibrational stability of this phase at finite temperatures 
as well as to evaluate its ionic transport properties. All the results presented in what follows, if not stated otherwise, 
were obtained with the semilocal PBEsol exchange-correlation energy functional \cite{pbesol}.

Figure~\ref{fig5} shows the position correlation function, $p(\Delta t)$, and mean square displacement, $\Delta r
(\Delta t)$, estimated for Ag$_{3}$SBr at $T = 200$ and $600$~K. These time-dependent functions are defined as 
\cite{cazorla12,cazorla14}:
\begin{eqnarray}
	p (\Delta t) = \langle [ {\bf r_{i}}(\Delta t + t_{0}) - {\bf R_{i}}^{0} ] \cdot [ {\bf r_{i}}(t_{0}) 
        - {\bf R_{i}}^{0} ] \rangle  \\
        \Delta r^{2} (\Delta t) = \langle [ {\bf r_{i}}(\Delta t + t_{0}) - {\bf r_{i}}(t_{0}) ]^{2} \rangle~,
\label{eq:pandmsd}
\end{eqnarray}
where ${\bf r_{i}}$ represents the position vector of atom $i$, $t_{0}$ an arbitrary time origin, ${\bf R_{i}}^{0}$ 
the position vector of the equilibrium lattice site for atom $i$, and $\langle \cdots \rangle$ thermal average 
performed over particles and time origins. 

At $\Delta t = 0$, $p$ is simply the vibrational mean square displacement. The crystal is vibrationally stable 
if $p (\Delta t \to \infty) = 0$, because the vibrational displacements at widely separated times become uncorrelated. 
Conversely, if the atoms acquire a permanent vibrational displacement, $p (\Delta t \to \infty)$ becomes nonzero. 
On the other hand, in the absence of ionic diffusion, $\Delta r^{2}$ converges to a constant equal to twice the 
vibrational mean square displacement in the limit $\Delta t \to \infty$. Contrarily, in the presence of ionic diffusion, 
$\Delta r^{2}$ exhibits a linear dependence on $\Delta t$ with a positively defined slope at sufficiently long times. 

The $p$ and $\Delta r^{2}$ results presented in Fig.\ref{fig5} indicate that Ag$_{3}$SBr in the cubic $Pm\overline{3}m$ 
phase is vibrationally stable at $T = 200$~K and exhibits significant Ag ionic diffusion at $T = 600$~K. The slow decay 
of $p$ and large asymptotic value of $\Delta r^{2}$ estimated for Ag ions at $T = 200$~K indicate high anharmonicity, 
despite of the low temperature. In contrast, the $p$ and $\Delta r^{2}$ results presented in Fig.\ref{fig6} indicate a 
markedly different behavior for Ag$_{3}$SI. Specifically, the vibrational centers of Ag ions somewhat change throughout 
the $T = 200$~K simulation, $\lbrace {\bf R_{i}}^{0} \rbrace \to \lbrace {\bf R_{i}}^{*} \rbrace$, since $p (\Delta t \to 
\infty) \neq 0$ for these atoms. Furthermore at high temperatures, the Ag ionic diffusion is appreciably higher in Ag$_{3}$SI 
than in Ag$_{3}$SBr.

Figure~\ref{fig7} shows the simulated trajectory of an arbitrary silver ion in Ag$_{3}$SBr and Ag$_{3}$SI at different 
temperatures, considering the cubic $Pm\overline{3}m$ phase. At $T = 200$~K, the map of position points is more anisotropic 
for Ag$_{3}$SI than for Ag$_{3}$SBr. Specifically, multiple vibrational centers can be identified for Ag$_{3}$SI 
(Fig.\ref{fig7}b), whereas they are not observed for Ag$_{3}$SBr (Fig.\ref{fig7}a). Notably, the amplitude of the ionic 
vibrations in Ag$_{3}$SBr is very wide, thus denoting high anharmonicity. These observations are consistent with the $p$ 
results presented in Figs.\ref{fig5}--\ref{fig6} for Ag ions. At $T = 600$~K, on the other hand, silver ionic hoppings are 
clearly observed in both compounds, illustrating their superionic character at high temperatures.  

In an attempt to identify the crystal structure towards which Ag$_{3}$SI appears to transform from the cubic $Pm\overline{3}m$ 
phase, we performed annealing (i.e., zero-temperature geometry relaxation) on a series of supercell configurations extracted 
from the AIMD runs (Fig.\ref{fig8}). Figure~\ref{fig8}b shows the lowest-energy configuration obtained through this process, 
which contains a total of $320$ atoms. Using the FINDSYM software \cite{findsym}, this supercell was efficiently reduced to a 
cubic $40$-atom primitive cell with space group $I2_{1}3$ (Fig.\ref{fig8}c). Interestingly, this new cubic $I2_{1}3$ phase 
was found to be vibrationally stable at low temperatures for both Ag$_{3}$SBr and Ag$_{3}$SI (Figs.\ref{fig3}--\ref{fig4}). 
As discussed in the next section, this new cubic $I2_{1}3$ phase was also found to be energetically competitive with respect 
to the ground-state phases determined for most silver chalcohalide anti-perovskite.

\begin{table*}
    \centering
    \begin{tabular}{ccccccccccc}
    \hline
    \hline
    & & & & & & & & & & \\
	    Compound & Structure & Symmetry& $a$~(\AA) & $b$~(\AA) & $c$~(\AA) & $\alpha$~$(^{\circ})$ & $\beta$~$(^{\circ})$ & $\gamma$~$(^{\circ})$ & $\Delta E$~(meV/f.u.) & $\Delta E^{\rm{ZPE}}$~(meV/f.u.)\\
    & & & & & & & & & & \\
    \hline
    \hline
    & & & & & & & & & & \\
	    \multirow{6}{*}{Ag$_{3}$SCl} & Cubic & $P2_13$~(I) & 7.478 & 7.478 & 7.478 & 90.000 & 90.000 & 90.000 & 0 & 0 \\
    & Orthorhombic & $P2_12_12_1$ & 4.502 & 4.455 & 20.715 & 90.000 & 90.000 & 90.000 & 65 & 64 \\
    & Orthorhombic & $Pca2_1$ & 5.872 & 9.757 & 6.877 & 90.000 & 90.000 & 90.000 & 79 & 78 \\
    & Cubic & $P2_13$~(II) & 7.499 & 7.499 & 7.499 & 90.000 & 90.000 & 90.000 & 89 & 89 \\
    & Cubic & $I2_13$ & 9.389 & 9.389 & 9.389 & 90.000 & 90.000 & 90.000 & 181 & -- \\
    & Cubic & $Pm\overline{3}m$ & 4.750 & 4.750 & 4.750 & 90.000 & 90.000 & 90.000 & 627 & -- \\
    & & & & & & & & & & \\
    \hline
    & & & & & & & & & & \\
	    \multirow{6}{*}{Ag$_{3}$SBr} & Cubic & $P2_13$~(I) & 7.586 & 7.586 & 7.586 & 90.000 & 90.000 & 90.000 & 0 & 0 \\
    & Orthorhombic & $P2_12_12_1$ & 4.518 & 4.531 & 21.152 & 90.000 & 90.000 & 90.000 & 79 & 77 \\
    & Cubic & $P2_13$~(II) & 7.602 & 7.602 & 7.602 & 90.000 & 90.000 & 90.000 & 91 & 90 \\
    & Cubic & $I2_13$ & 9.446 & 9.446 & 9.446 & 90.000 & 90.000 & 90.000 & 117 & 109 \\
    & Trigonal & $R3$ & 8.140 & 8.140 & 8.140 & 109.075 & 109.075 & 109.075 & 121 & 115 \\
    & Cubic & $Pm\overline{3}m$ & 4.793 & 4.793 & 4.793 & 90.000 & 90.000 & 90.000 & 538 & -- \\
    & & & & & & & & & & \\
    \hline
    & & & & & & & & & & \\
	    \multirow{6}{*}{Ag$_{3}$SI} & Cubic & $P2_13$~(I) & 7.734 & 7.734 & 7.734 & 90.000 & 90.000 & 90.000 & 0 & 0 \\
    & Cubic & $I2_13$ & 9.568 & 9.568 & 9.568 & 90.000 & 90.000 & 90.000 & 53 & 47 \\
    & Monoclinic & $P2_1$ & 6.176 & 6.984 & 5.656 & 90.000 & 90.000 & 93.869 & 55 & 55 \\
    & Cubic & $P2_13$~(II) & 7.739 & 7.739 & 7.739 & 90.000 & 90.000 & 90.000 & 67 & 73 \\
    & Orthorhombic & $P2_12_12_1$ & 4.563 & 4.612 & 21.767 & 90.000 & 90.000 & 90.000 & 79 & 77 \\
    & Cubic & $Pm\overline{3}m$ & 4.869 & 4.869 & 4.869 & 90.000 & 90.000 & 90.000 & 516 & -- \\
    & & & & & & & & & & \\
    \hline
    & & & & & & & & & & \\
	    \multirow{6}{*}{Ag$_{3}$SeCl} & Orthorhombic & $Pca2_1$ & 5.879 & 10.014 & 6.980 & 90.000 & 90.000 & 90.000 & 0 & 0 \\
    & Orthorhombic & $P2_12_12_1$~(I) & 4.688 & 4.475 & 21.053 & 90.000 & 90.000 & 90.000 & 4 & 4 \\
    & Orthorhombic & $P2_12_12_1$~(II) & 5.761 & 5.743 & 12.651 & 90.000 & 90.000 & 90.000 & 38 & 39 \\
    & Orthorhombic & $Pna2_1$ & 6.347 & 6.450 & 9.966 & 90.000 & 90.000 & 90.000 & 43 & 38 \\
    & Cubic & $I2_13$ & 9.626 & 9.626 & 9.626 & 90.000 & 90.000 & 90.000 & 117 & -- \\
    & Cubic & $Pm\overline{3}m$ & 4.930 & 4.930 & 4.930 & 90.000 & 90.000 & 90.000 & 1072 & -- \\
    & & & & & & & & & & \\
    \hline
    & & & & & & & & & & \\
\multirow{6}{*}{Ag$_{3}$SeBr} & Orthorhombic & $P2_12_12_1$~(I) & 4.706 & 4.562 & 21.452 & 90.000 & 90.000 & 90.000 & 0 & 0 \\
    & Orthorhombic & $P2_12_12$~(II) & 7.530 & 14.614 & 4.365 & 90.000 & 90.000 & 90.000 & 10 & 9 \\
    & Cubic & $I2_13$ & 9.678 & 9.678 & 9.678 & 90.000 & 90.000 & 90.000 & 41 & 37 \\
    & Cubic & $P2_13$~(I) & 7.739 & 7.739 & 7.739 & 90.000 & 90.000 & 90.000 & 47 & 46 \\
    & Cubic & $P2_13$~(II) & 7.737 & 7.737 & 7.737 & 90.000 & 90.000 & 90.000 & 49 & 48 \\
    & Cubic & $Pm\overline{3}m$ & 4.962 & 4.962 & 4.962 & 90.000 & 90.000 & 90.000 & 916 & -- \\
    & & & & & & & & & & \\
    \hline
    & & & & & & & & & & \\
	    \multirow{6}{*}{Ag$_{3}$SeI} & Cubic & $I2_13$ & 9.782 & 9.782 & 9.782 & 90.000 & 90.000 & 90.000 & 0 & 0 \\
    & Orthorhombic & $P2_12_12_1$ & 4.761 & 4.653 & 21.970 & 90.000 & 90.000 & 90.000 & 19 & 22 \\
    & Trigonal & $P3_1$ & 6.949 & 6.949 & 8.388 & 90.000 & 90.000 & 120.000 & 52 & 51 \\
    & Monoclinic & $P2_1$ & 7.708 & 7.225 & 15.533 & 90.000 & 90.000 & 144.418 & 66 & 70 \\
    & Cubic & $P2_13$~(I) & 7.978 & 7.978 & 7.978 & 90.000 & 90.000 & 90.000 & 69 & 68\\
    & Cubic & $Pm\overline{3}m$ & 5.022 & 5.022 & 5.022 & 90.000 & 90.000 & 90.000 & 842 & -- \\
    & & & & & & & & & & \\
    \hline
    \hline
    \end{tabular}
    \caption{\textbf{Structural and energy properties of CAP considering several energetically competitive phases.} 
    Energy differences, $\Delta E$, are referred to the ground-state phase and expressed in units of meV per formula unit 
    (f.u.). Energy differences including zero-point energy (ZPE) corrections (Methods) are also reported, $\Delta E^{\rm ZPE}$, 
    except for those phases that were found to exhibit imaginary phonon frequencies. Results were obtained with the 
    semi-local functional PBEsol \cite{pbesol}.}
    \label{tab1}
    \end{table*}

\begin{figure*}
    \centering
    \includegraphics[width=\textwidth]{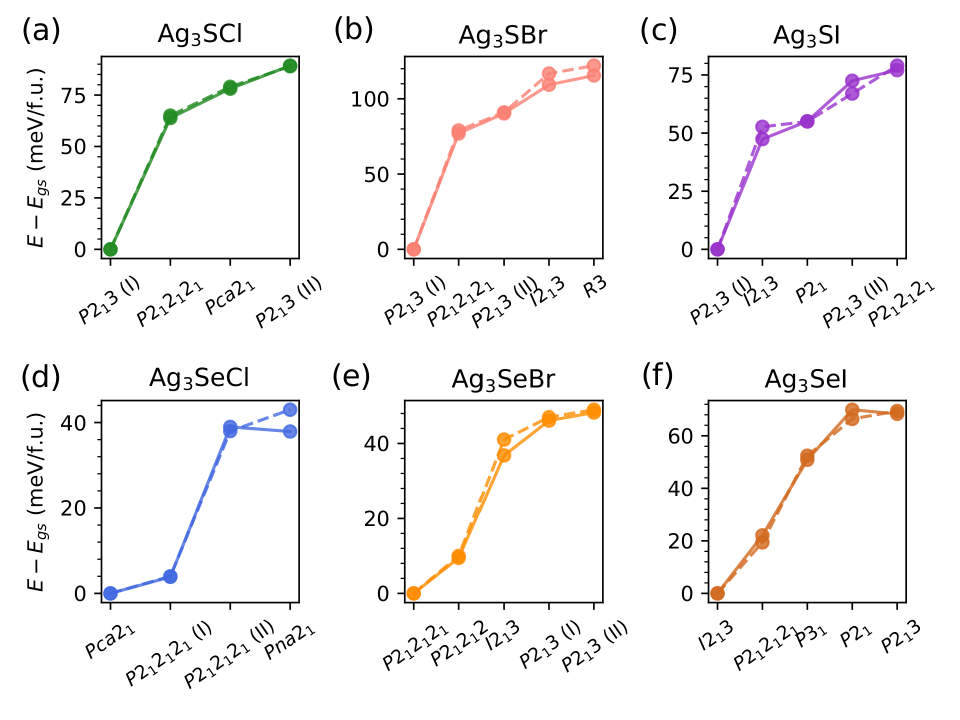}
	\caption{\textbf{CAP phase competition determined at zero-temperature conditions.} Energy differences neglecting 
	(including) ZPE corrections (Methods) are represented with solid (dashed) lines. Results were obtained with the 
	semi-local functional PBEsol \cite{pbesol}.}
    \label{fig9}
\end{figure*}

\subsection{Zero-temperature phase competition}
\label{subsec:zeropd}
After the detailed study of the structural and vibrational properties of Ag$_{3}$SBr and Ag$_{3}$SI, we extended our
investigation to other materials in the CAP family, namely, Ag$_{3}$SCl, Ag$_{3}$SeCl, Ag$_{3}$SeBr and Ag$_{3}$SeI.
It is worth noting that, to the best of our knowledge, there are not experimental reports on these compounds in the 
literature. For each new CAP, we began by relaxing the $60$ lowest-energy phases identified for the archetypal compounds 
Ag$_{3}$SBr and Ag$_{3}$SI, using the semilocal PBEsol exchange-correlation energy functional \cite{pbesol}. The cubic 
$I2_{1}3$ phase discovered from the AIMD simulations performed for Ag$_{3}$SI and the experimental room-temperature 
cubic $Pm\overline{3}m$ phase, were additionally considered in our pool of candidate ground-state phases.

Figure~\ref{fig9} and Table~\ref{tab1} summarize our zero-temperature DFT structure energy rankings obtained for the 
CAP family. For Ag$_{3}$SBr and Ag$_{3}$SI, the primary difference compared to the results in Fig.\ref{fig2} is the 
inclusion of the new cubic $I2_{1}3$ phase. This phase has an energy only $45$~meV/f.u. higher than the cubic 
$P2_{1}3$~(I) ground-state phase in Ag$_{3}$SI, and approximately $100$~meV/f.u. higher than the same ground-state 
phase in Ag$_{3}$SBr. For the analogous Se-based CAP, the cubic $I2_{1}3$ phase is even more relevant; it has the 
lowest energy for Ag$_{3}$SeI, and for Ag$_{3}$SeBr its energy is only about $35$~meV/f.u. higher than that of the 
orthorhombic $P2_{1}2_{1}2_{1}$ ground-state phase (Fig.\ref{fig9}). 

The ground-state phase of all analyzed S-based CAP corresponds to the cubic $P2_{1}3$~(I) phase, which in the 
Ag$_{3}$SCl system is closely followed by two orthorhombic phases with crystal symmetries $P2_{1}2_{1}2_{1}$ and 
$Pca2_{1}$. Regarding the Se-based CAP, in Ag$_{3}$SeCl and Ag$_{3}$SeBr, the lowest-energy phases are dominated 
by orthorhombic phases (e.g., $P2_{1}2_{1}2_{1}$ and $Pca2_{1}$), the cubic $I2_{1}3$ phase being the most significant 
energy competitor presenting a different crystal symmetry. In the specific case of Ag$_{3}$SeI, a trigonal $P3_{1}$ 
and a monoclinic $P2_{1}$ phase were also ranked among the lowest-energy structures.     

We computed the lattice phonon spectrum of the experimental room-temperature cubic $Pm\overline{3}m$ phase and the
five energetically most favourable structures found for Ag$_{3}$SCl, Ag$_{3}$SeCl, Ag$_{3}$SeBr and Ag$_{3}$SeI 
(Supplementary Figs.4--7). Without any exception, we found that the cubic $Pm\overline{3}m$ phase always exhibits 
abundant imaginary phonon frequencies. The cubic $I2_{1}3$ phase is also vibrationally unstable for Ag$_{3}$SCl and 
Ag$_{3}$SeCl at low temperatures (Supplementary Figs.4--5). However, the remaining phases are perfectly vibrationally 
stable for all the targeted CAP (Supplementary Figs.4--7). 

For the structures that exclusively exhibited real and positively defined phonon frequencies, we corrected their 
zero-temperature energies to account for likely quantum zero-point effects (ZPE) \cite{cazorla17,cazorla09,cazorla22}. 
It is worth noting that applying quasi-harmonic (QH) approaches to materials that exhibit imaginary phonon frequencies 
is neither physically nor mathematically well-justified \cite{cazorla17}. Thus, any QH result obtained for phases 
that exhibit imaginary phonon frequencies (e.g., the cubic $Pm\overline{3}m$ phase for all CAP compounds) cannot be 
rigorously trusted as quantitatively correct \cite{yin20}. As shown in Fig.\ref{fig9} and Table~\ref{tab1}, accounting 
for quantum ZPE has a practically negligible effect on the energy rankings obtained using classical mechanics. Only 
in the specific cases of Ag$_{3}$SeCl and Ag$_{3}$SeI, quantum ZPE corrections vary the ordering between the third 
and fourth and the fourth and fifth most energetically favourable phases, respectively.

\begin{figure*}
    \centering
    \includegraphics[width=\textwidth]{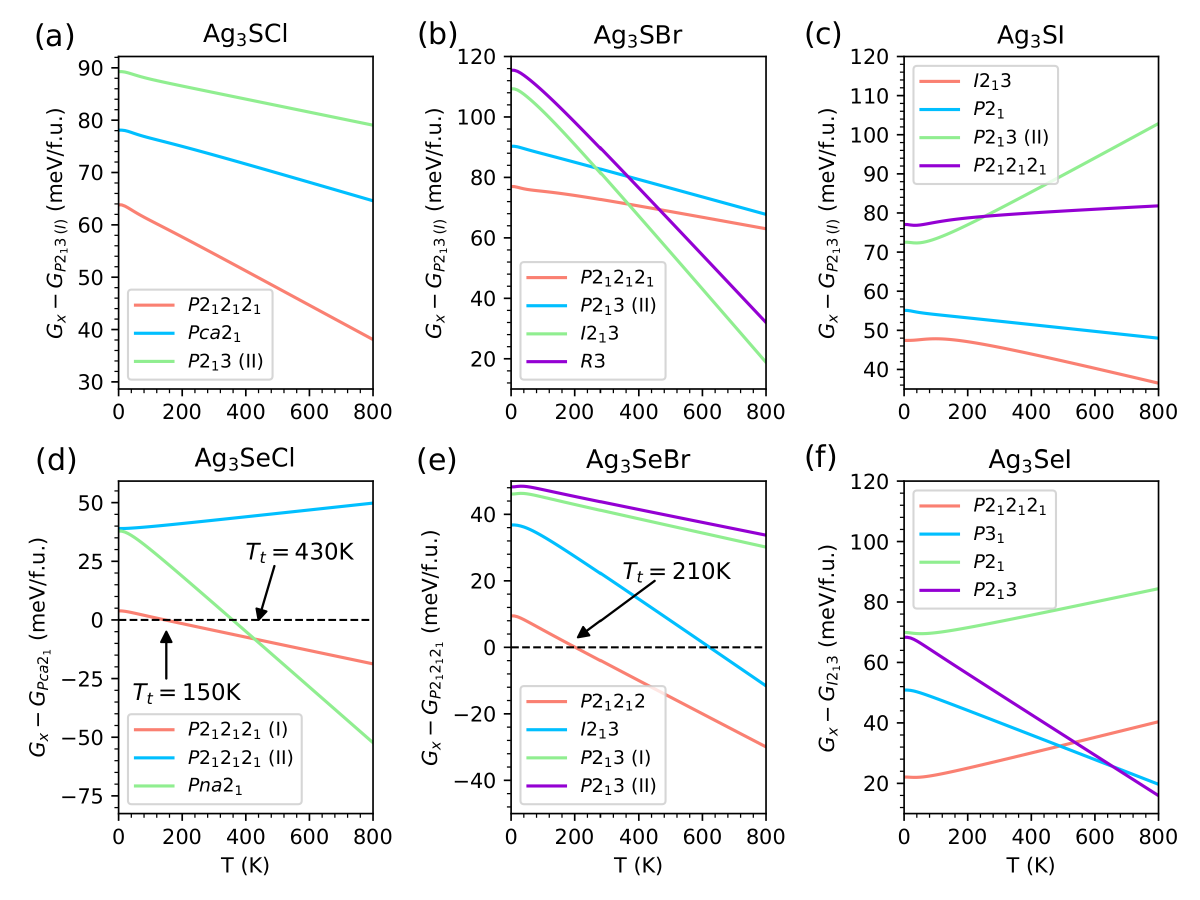}
	\caption{\textbf{CAP phase competition determined at $T \neq 0$ conditions.} QH Helmholtz free-energy differences, 
	$\Delta G$, were calculated within the quasi-harmonic approximation (Methods) \cite{cazorla17,cazorla09,cazorla22}. 
	Results were obtained with the semi-local functional PBEsol \cite{pbesol}.}
    \label{fig10}
\end{figure*}

\begin{figure*}[t]
    \centering
    \includegraphics[width=\linewidth]{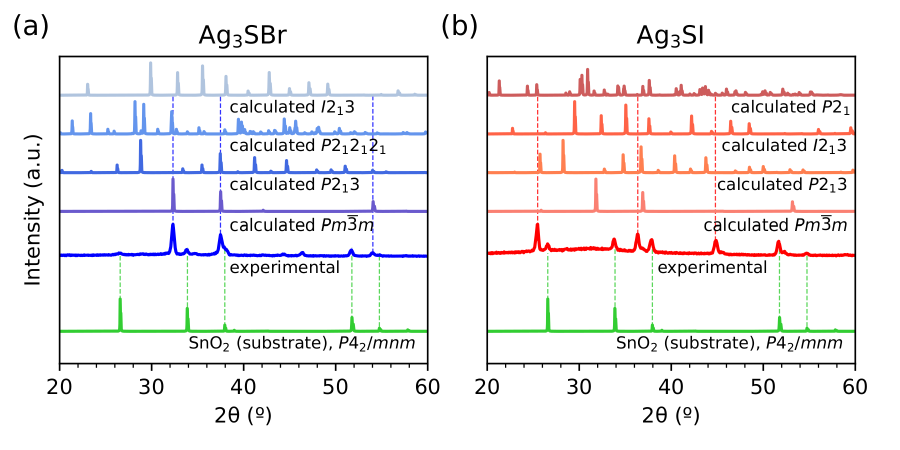}
        \caption{\textbf{Experimental X-ray characterization of CAP at room temperature.} (a)~Ag$_{3}$SBr and
        (b)~Ag$_{3}$SI. Theoretical DFT results obtained for different energetically competitive phases are shown
        for comparison.}
    \label{fig11}
\end{figure*}

\subsection{Finite-temperature phase competition}
\label{subsec:finitepd}
To predict likely $T$-induced phase transitions across the CAP family, we applied the QH free-energy formalism 
\cite{cazorla17,cazorla09,cazorla22} to all the phases reported in Fig.\ref{fig9} that were vibrationally stable 
at low temperatures. It is worth emphasizing that applying QH approaches to phases that exhibit imaginary phonon 
frequencies is neither physically nor mathematically well-justified \cite{cazorla17}. Consequently, phases like 
the cubic $Pm\overline{3}m$ and orthorhombic $Cmcm$ in Ag$_{3}$SBr, for example, were excluded from our initial
phase competition analysis conducted at $T \neq 0$~K conditions (although the thermodynamic stability of some 
of these phases were subsequently assessed by considering $T$-renormalized phonons, see next Sec.~\ref{sec:discussion}).  

Figure~\ref{fig10} shows the QH (Helmholtz) free-energy differences estimated as a function of temperature for the 
six CAP compounds targeted in this study. At a given temperature, the phase exhibiting the lowest free energy 
($G$) is the equilibrium, or stable, phase. The remaining phases, which are vibrationally stable in principle, 
are considered metastable. A $T$-induced phase transition occurs when the free-energy curve of the equilibrium 
phase crosses the free-energy curve of a metastable phase. The temperature at which this crossing occurs 
defines the corresponding phase transition temperature. 

For the S-based CAP, we did not find any $T$-induced phase transition up to $800$~K as evidenced by the lack 
of $G$ curve crossings involving the cubic $P2_{1}3$~(I) phase. Considering higher temperatures probably is not 
physically meaningful due to the imminent stabilization of the liquid phase \cite{luna23,cano24}. Several $G$ 
curve crossings involving pairs of metastable phases are observed in Figs.\ref{fig10}b--c; however, such 
metastable phase transitions are not of interest in this study. Consequently, according to our QH free-energy 
DFT calculations, the cubic $P2_{1}3$~(I) phase is the equilibrium, or stable, phase of Ag$_{3}$SCl, Ag$_{3}$SBr 
and Ag$_{3}$SI at room temperature (the same conclusion is obtained when considering $T$-renormalized phonons 
for the cubic $Pm\overline{3}m$ phase, see next Sec.~\ref{sec:discussion}). 

For the Se-based CAP, the equilibrium between different phases is appreciably affected by temperature. Specifically, 
we predict two phase transitions for Ag$_{3}$SeCl: the first occurrs between the two orthorhombic phases $Pca2_{1}$ 
and $P2_{1}2_{1}2_{1}$~(I) at $150$~K, and the second between the two orthorhombic phases $P2_{1}2_{1}2_{1}$~(I) and 
$Pna2_{1}$ at $430$~K (Fig.\ref{fig10}d). For Ag$_{3}$SeBr, another phase transition is predicted to occur between 
the two orthorhombic phases $P2_{1}2_{1}2_{1}$~(I) and $P2_{1}2_{1}2_{1}$~(II) at $210$~K (Fig.\ref{fig10}e). Finally, 
for Ag$_{3}$SeI no $T$-induced phase transition is observed (Fig.\ref{fig10}f). Consequently, according to our 
QH free-energy DFT calculations, the equilibrium room-temperature phase is orthorhombic $P2_{1}2_{1}2_{1}$ for
Ag$_{3}$SeCl and Ag$_{3}$SeBr, and cubic $I2_{1}3$ for Ag$_{3}$SeI.

\begin{figure*}[t]
    \centering
    \includegraphics[width=\linewidth]{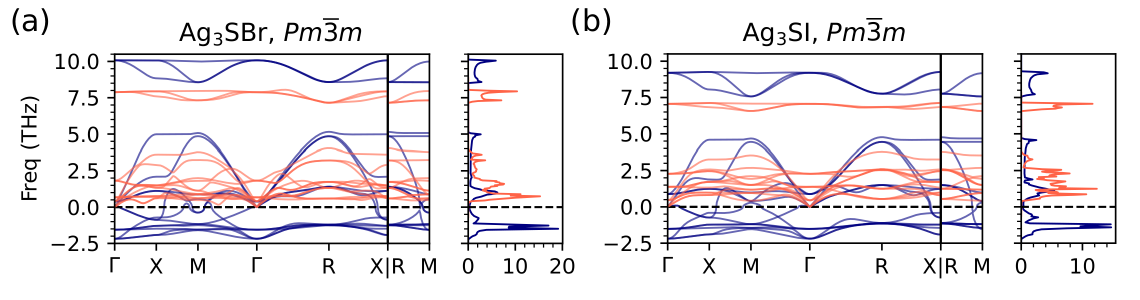}
	\caption{\textbf{Temperature-renormalized vibrational phonon spectrum of CAP in the cubic $Pm\overline{3}m$ phase.} 
	(a)~Ag$_{3}$SBr and (b)~Ag$_{3}$SI. Phonon frequencies obtained at $T = 100$~K (Methods) are represented with orange 
	lines. Zero-temperature phonon frequencies are represented with blue lines. Results were obtained with the semi-local 
	functional PBEsol \cite{pbesol}.}
    \label{fig12}
\end{figure*}

\section{Discussion}
\label{sec:discussion}
Figure~\ref{fig11} shows the experimental X-ray diffraction patterns obtained for two Ag$_{3}$SBr and Ag$_{3}$SI 
polycrystalline samples synthesized through the chemical route detailed in work \cite{cano24} (Methods). At room 
temperature, using LeBail phase identification analysis (Methods), the Ag$_{3}$SBr film was indexed as cubic 
$Pm\overline{3}m$ and the Ag$_{3}$SI film as cubic $Im\overline{3}m$ (likely due to the presence of ionic disorder). 
In the same plot, the theoretical X-ray diffraction patterns of several phases obtained from our DFT simulations 
are included for comparison. The experimental X-ray diffraction pattern of the Ag$_{3}$SBr sample perfectly agrees 
with the theoretical X-ray diffraction pattern of the simulated cubic $Pm\overline{3}m$ phase (Fig.\ref{fig11}a). 
On the other hand, the experimental X-ray diffraction pattern of the Ag$_{3}$SI sample cannot be assigned to the 
theoretical pattern of any single ordered structure (Fig.\ref{fig11}b).

The experimental X-ray diffraction patterns enclosed in Fig.\ref{fig11}, along with the computational \emph{ab initio} 
findings presented in the preceding sections, raise the following question: why is the cubic $Pm\overline{3}m$ 
phase, or the similar one $Im\overline{3}m$, experimentally identified for Ag$_{3}$SBr and Ag$_{3}$SI at room temperature, 
instead of the theoretically predicted equilibrium cubic $P2_{1}3$ phase? Although we cannot provide a definitive 
answer to this question, we can offer some reasonable hypotheses. 

First, it is noted that despite the cubic $Pm\overline{3}m$ phase being vibrationally unstable at very low temperatures
(Figs.\ref{fig3}--\ref{fig4}), it could be stabilized under increasing temperature, similar to what is observed in 
archetypal oxide perovskites \cite{ehsan22,cazorla24,cazorla24b}. To evaluate this possibility, we computed the $T$-renormalized 
phonon spectrum of this phase for Ag$_{3}$SBr and Ag$_{3}$SI following a dynamical normal-mode-decomposition technique 
(Methods) \cite{dynaphopy}. Using a relatively small $2 \times 2 \times 2$ supercell and a dense {\bf k}-point grid 
for the AIMD simulations ($T=100$~K) involved in these calculations, to maintain high accuracy and numerical 
consistency (Supplementary Discussion), we found that, in fact, the cubic $Pm\overline{3}m$ phase is vibrationally 
stabilized under increasing temperature (Fig.\ref{fig12}).  

Importantly, since no imaginary phonon frequencies appear in the $T$-renormalized phonon spectra of Ag$_{3}$SBr
and Ag$_{3}$SI (Fig.\ref{fig12}) we can estimate the QH free energy of the cubic $Pm\overline{3}m$ phase and 
compare it with those of other phases. Our $T$-renormalized phonon QH free-energy calculations (Supplementary Fig.8) 
show that the cubic $Pm\overline{3}m$ phase is not thermodynamically stabilized over the other considered phases at 
temperature $T \le 800$~K, that is, its free energy is never a minimum, in spite of its rapid $G$ decrease observed 
under increasing temperature. For example, at zero temperature the free-energy difference between the cubic $Pm\overline{3}m$ 
and ground-state phases approximately amounts to $0.5$~eV/f.u. for Ag$_{3}$SBr, while at $T = 800$~K this free-energy 
difference is reduced to $0.1$~eV per formula unit.

Based on the presented theoretical findings, our main hypothesis is that current synthesis methods, all of 
which involve temperatures significantly higher than $300$~K, inevitably trap CAP compounds into metastable 
states. A key feature, possibly explaining the room-temperature metastability of the synthesized CAP compounds, 
may be superionicity. According to our AIMD simulations (Supplementary Figs.9--10), Ag$_{3}$SBr in the cubic 
$Pm\overline{3}m$ phase becomes superionic at $T = 300~(100)$~K. On other hand, the equibrium cubic $P2_{1}3$ 
phase exhibits no ionic diffusivity at room temperature. Consequently, it is likely that at temperatures 
higher than $300$~K, such as those involved in CAP synthesis methods, the cubic $Pm\overline{3}m$ phase is 
further stabilized due to the entropy gain that accompanies ionic diffusion, eventually becoming the equilibrium 
phase. (Unfortunately, this type of entropy enhancement resulting from ionic disorder cannot be reproduced by 
our QH free-energy DFT calculations.) In that hypothetical case, it seems plausible that upon annealing the 
system becomes trapped in the cubic $Pm\overline{3}m$ phase with little possibility to nucleate the equilibrium 
ordered phase corresponding to ambient temperature.

\section{Conclusions}
\label{sec:conclusions}
For the highly anharmonic and archetypal CAP compounds Ag$_{3}$SBr and Ag$_{3}$SI, our systematic, comprehensive 
and technically sound first-principles study suggests a completely different set of stable phases, both at zero 
temperature and $T \neq 0$~K conditions, compared to what has been proposed from experiments. Specifically, a new 
cubic $P2_{1}3$ phase is found to exhibit the lowest energy at $T = 0$~K and the lowest free energy at finite 
temperatures, being several hundreds of meV per formula unit lower than the experimentally proposed trigonal 
$R3$ and orthorhombic $Cmcm$ phases. Additionally, we propose different ground-state phases and $T$-induced 
phase transitions for the less investigated CAP compounds Ag$_{3}$SCl, Ag$_{3}$SeCl, Ag$_{3}$SeBr and Ag$_{3}$SeI.

An interesting consequence of our theoretical findings is that the cubic $Pm\overline{3}m$ phase experimentally 
identified at room temperature may be metastable. The relatively high temperatures involved in current synthesis 
methods and the significant superionicity of CAP above room temperature may be key factors in understanding this 
metastability. It is worth noting that metastable phases have a natural tendency to fluctuate between different 
energy states, which may pose stability issues for practical technological applications utilizing CAP. Thus, the 
present computational study calls for a revision of the prevalent phase diagrams of CAP compounds and motivates 
new structural characterization experiments conducted under varying temperature conditions, particularly in the 
low-temperature regime where ionic disorder and metastability can be minimized.

\section*{Methods}
\label{sec:methods}
{\bf First-principles calculations outline.}~\textit{Ab initio} calculations based on density functional theory (DFT) 
\cite{cazorla17} were performed to analyse the structural and phase stability properties of CAP. We performed 
these calculations with the VASP code \cite{vasp} using different approximations to the exchange-correlation energy, namely, 
the local and semi-local functionals LDA \cite{lda}, PBE \cite{pbe}, PBEsol \cite{pbesol} and SCAN \cite{scan}. The 
projector augmented-wave method was used to represent the ionic cores \cite{bloch94} and the following electronic states 
were considered as valence: Ag $5s$ $4d$, S $3s$ $3p$, Se $4s$ $4p$, Cl $3s$ $3p$, Br $4s$ $4p$, I $5s$ $5p$. Wave functions
were represented in a plane-wave basis typically truncated at $650$~eV. By using these parameters and dense {\bf k}-point 
grids for reciprocal-space integration (e.g., of $8 \times 8 \times 8$ for the cubic $5$-atom $Pm\overline{3}m$ unit cell), 
the resulting zero-temperature energies were converged to within $1$~meV per formula unit. In the geometry relaxations, a 
force tolerance of $0.005$~eV$\cdot$\AA$^{-1}$ was imposed in all the atoms.
\\

{\bf Harmonic and anharmonic phonon and free-energy calculations.}~The second-order interatomic force constant of all 
CAP and resulting harmonic phonon spectrum were calculated with the finite-differences method as is implemented in the 
PhonoPy software \cite{phonopy}. Large supercells (e.g., $4 \times 4 \times 4$ for the cubic $Pm\overline{3}m$ 
phase containing $320$ atoms) and a dense {\bf k}-point grid of $3 \times 3 \times 3$ for Brillouin zone (BZ) sampling 
were employed for the phonon calculations of targeted structures. Several numerical tests were conducted that demonstrated 
the adequacy of the selected {\bf k}-point grid (Supplementary Fig.11). Zero-point energy (ZPE) corrections and 
finite-temperature Helmholtz free energies ($G$) were calculated within the quasi-harmonic approximation 
\cite{cazorla17,cazorla09,cazorla22}. Due to the large number of materials and phases analyzed in this study, thermal 
expansion effects were disregarded in our ZPE and $G$ calculations. 

The DynaPhoPy code \cite{dynaphopy} was used to calculate the anharmonic lattice dynamics (i.e., $T$-renormalized 
phonons) of Ag$_{3}$SBr and Ag$_{3}$SI in the cubic $Pm\overline{3}m$ phase from \textit{ab initio} molecular dynamics 
(AIMD) simulations. (For the rest of phases, which do not exhibit imaginary phonon frequencies, anharmonic 
lattice dynamics effects were disregarded.) In this case, a reduced $2 \times 2 \times 2$ supercell and $4 \times 4 
\times 4$ {\bf k}-point grid for BZ sampling were employed in the AIMD simulations to maintain high numerical 
accuracy (Supplementary Discussion and Supplementary Fig.12). 

A normal-mode-decomposition technique was employed in which the atomic velocities $\textbf{v}_{jl}(t)$ ($j$ and $l$ 
represent particle and Cartesian direction indexes) generated during fixed-temperature AIMD simulation runs were 
expressed like:
\begin{equation}
\textbf{v}_{jl} (t) = \frac{1}{\sqrt{N m_{j}}} \sum_{\textbf{q}s}\textbf{e}_{j}(\textbf{q},s) 
	e^{i \textbf{q} \textbf{R}_{jl}^{0}} v_{\textbf{q}s}(t)~,
\label{eq1}
\end{equation}
where $N$ is the number of particles, $m_{j}$ the mass of particle $j$, $\textbf{e}_{j}(\textbf{q},s)$ a phonon mode eigenvector 
($\textbf{q}$ and $s$ stand for the wave vector and phonon branch), $\textbf{R}_{jl}^{0}$ the equilibrium position of particle $j$, 
and $v_{\textbf{q}s}$ the velocity of the corresponding phonon quasiparticle. 

The Fourier transform of the autocorrelation function of $v_{\textbf{q}s}$ then was calculated, yielding the power spectrum:
\begin{equation}
G_{\textbf{q}s} (\omega) = 2 \int_{-\infty}^{\infty} \langle v_{\textbf{q}s}^{*}(0) v_{\textbf{q}s}(t) \rangle e^{i \omega t} dt~. 
\label{eq2}
\end{equation}
Finally, this power spectrum was approximated by a Lorentzian function of the form:
\begin{equation}
G_{\textbf{q}s} (\omega) \approx \frac{\langle |v_{\textbf{q}s}|^{2} \rangle}{\frac{1}{2} \gamma_{\textbf{q}s} 
        \pi \left[ 1 + \left( \frac{\omega - \omega_{\textbf{q}s}}{\frac{1}{2}\gamma_{\textbf{q}s}} \right)^{2}  \right]}~, 
\label{eq3}
\end{equation}
from which a $T$-renormalized quasiparticle phonon frequency, $\omega_{\textbf{q}s} (T)$, was determined as the peak position, 
and the corresponding phonon linewidth, $\gamma_{\textbf{q}s} (T)$, as the full width at half maximum.
\\

{\bf First-principles molecular dynamics simulations.}~\emph{Ab initio} molecular dynamics simulations based on DFT were 
performed in the canonical $(N,V,T)$ ensemble (i.e., constant number of particles, volume and temperature). The selected 
volumes were those determined at zero temperature hence thermal expansion effects were neglected. The temperature in the AIMD 
simulations was kept fluctuating around a set-point value by using Nose-Hoover thermostats. Large simulation boxes containing 
$N \sim 200$--$400$ atoms were employed (e.g., for the cubic $Pm\overline{3}m$ phase we adopted a $4 \times 4 \times 4$ 
supercell contaning $320$ atoms) and periodic boundary conditions were applied along the three supercell lattice vectors. 
Newton's equations of motion were integrated using the customary Verlet's algorithm with a time step of $1.5 \cdot 10^{-3}$~ps. 
$\Gamma$-point sampling for reciprocal-space integration was employed in most of the AIMD simulations, which comprised long 
simulation times of $80$--$100$~ps.
\\

{\bf Crystal structure searches.}~We used the MAGUS software (Machine learning And Graph theory assisted Universal 
structure Searcher) \cite{magus} to find new candidate stable and metastable phases for the archetypal CAP Ag$_{3}$SBr 
and Ag$_{3}$SI. This crystal structure prediction software employs an evolutionary algorithm augmented with machine learning 
and graph theory to reduce the cost of the geometry optimizations. The crystal phase searches were conducted for structures 
containing a maximum of $4$ formula units (i.e., $20$ atoms per unit cell).   
\\

{\bf Experimental procedure.}~Ag$_{3}$SBr and Ag$_{3}$SI were synthesized following the chemical route described in work 
\cite{cano24}. The samples were characterized by X-ray diffraction using a Bruker D8 advanced diffractometer equipped with 
a Cu-based tube ($40$~kV, $40$~mA) and a Sol-X detector with discriminator for the $K_{\beta}$ line. The measurements were 
conducted using grazing incidence configuration and the data was analyzed with the LeBail profile fitting method as 
implemented in the FullProf suite software \cite{fullprof}. 
\\

\section*{Data availability}
The data that support the findings of this study are freely available at the NOMAD data management platform for 
materials science \cite{nomad1}. These data include the VASP input files of our first-principles DFT calculations 
as well as some key output files (e.g., relaxed atomic positions of all relevant crystal structures) \cite{nomad2}.
\\

\section*{Acknowledgements}
C.C. acknowledges support from the Spanish Ministry of Science, Innovation and Universities under the fellowship
RYC2018-024947-I and grants PID2020-112975GB-I00 and grant TED2021-130265B-C22. The authors also thankfully acknowledge
technical support the computational resources at MareNostrum4 provided by Barcelona Supercomputing Center (FI-2023-1-0002, 
FI-2023-2-0004, FI-2023-3-0004 and FI-2024-1-0005). P.B. acknowledges support from the Generalitat de Catalunya under 
a FI grant. C.L. acknowledges support from the Spanish Ministry of Science, Innovation and Universities under a FPU grant.
This publication and other research outcomes are supported by the predoctoral program AGAUR-FI ajuts (2024 FI-3 00065) 
Joan Oró, which is backed by the Secretariat of Universities and Research of the Department of Research and Universities 
of the Generalitat of Catalonia, as well as the European Social Plus Fund. This work is part of Maria de Maeztu Units of 
Excellence Programme CEX2023-001300-M.
\\

\end{document}